# Shortest path or random walks? A framework for path weights in network meta-analysis

Gerta Rücker*[1] | Theodoros Papakonstantinou[1] | Adriani Nikolakopoulou[1] | Guido Schwarzer[1] | Tobias Galla[2] | Annabel L. Davies[3]

[1]Institute of Medical Biometry and Statistics, Medical Faculty and Medical Center - University of Freiburg, Freiburg, Germany

[2]Instituto de Física Interdisciplinar y Sistemas Complejos, IFISC (CSIC-UIB), Campus Universitat Illes Balears, Palma de Mallorca, Spain

[3]Bristol Medical School, University of Bristol, Bristol, United Kingdom

**Correspondence**
*Email: gerta.ruecker@uniklinik-freiburg.de

**Present Address**
Stefan-Meier-Strasse 26, D-79104 Freiburg, Germany

**Summary**

Quantifying the contributions, or weights, of comparisons or single studies to the estimates in a network meta-analysis (NMA) is an active area of research. We extend this to the contributions of paths to NMA estimates. We present a general framework, based on the path-design matrix, that describes the problem of finding path contributions as a linear equation. The resulting solutions may have negative coefficients. We show that two known approaches, called shortestpath and randomwalk, are special solutions of this equation, and both meet an optimization criterion, as they minimize the sum of absolute path contributions. In general, there is an infinite space of solutions, which can be identified using the generalized inverse (Moore-Penrose pseudoinverse). We consider two further special approaches. For complex networks we find that shortestpath is superior with respect to run time and variability, compared to the other approaches, and is thus recommended in practice. The path-weights framework also has the potential to answer more general research questions in network meta-analysis.

## 1 | INTRODUCTION

Network meta-analysis (NMA) is a generalization of pairwise meta-analysis to compare effects of multiple treatments, based on controlled studies. Under the assumption of transitivity, NMA allows one to estimate treatment effects for all comparisons in the network, even if not directly observed. NMA makes use of indirect evidence, meaning that a primary study can contribute information about the effects of treatments that have not been compared in that study. Quantifying the contributions of comparisons or single studies to the estimates in an NMA has been an active area of research for some time.[1,2,3,4,5,6]

In pairwise meta-analysis, where the summary effect is a simple weighted average of the study effects, one can naturally use the study weights to define contributions. The generalization of the concept to NMA has proven not to be trivial and multiple methods were proposed. This was initially motivated by the concern that some comparisons or studies may be associated with high risk of bias, which may heavily contribute to some, or even all, network estimates. Therefore the aim was to quantify the contribution, or the weight, of each comparison or study to each NMA estimate.

The first approach to this aim in the framework of a frequentist NMA approach was made in a paper by Salanti and others.[1] It was based on normalizing the rows of the hat matrix which maps the observed treatment effect estimates to their NMA counterparts. Later, the authors of this paper detected that this approach has limitations and may lead to erroneous and inconsistent contributions. This led to a new proposal to derive the proportion (percentage) contributions (edge weights). Based on seminal work by König et al.[7], Papakonstantinou et al. introduced the idea of paths of evidence and developed the shortest path algorithm, a greedy algorithm which iteratively assigns contributions to each path, removes the path from the network and iterates this process until the network is disconnected.[4] This approach is currently implemented in CINeMA (Confidence in Network Meta-Analysis), a web application to simplify the evaluation of confidence in the findings from network meta-analysis.[8,9] The authors noted that contributions derived in this way could also start from a random path, or from the longest path, and the results would not always be the same, thus leaving some ambiguity.[4] The same group performed an empirical analysis of the contributions of paths in published networks.[6]

Recently, Davies et al. continued this work by using a novel analogy between NMA and random walks which can be used to derive closed-form expressions for the path contributions that are mathematically sound.[10] By introducing the idea of looking for all paths that join two treatment nodes in a network, this work opened a new road to investigating the flow of evidence through a network. Both the shortest path (*shortestpath*) and random walk (*randomwalk*) algorithms are implemented in function `netcontrib` of R package **netmeta** for frequentist NMA.[11] For details of these two approaches, we refer to the original articles.[4,10]

In this paper we will mainly use the more concise term 'weight' (instead of contribution) and distinguish between path weights and edge weights. We first describe a real data example in Section 2. In Section 3 we introduce paths, path weights and various path-based approaches. We start by introducing our notation and define several matrices, such as the full hat matrix (Subsection 3.1) and the path-design matrix (Subsection 3.2), demonstrating all notions on two small fictitious examples. In Subsection 3.3 we demonstrate that path weights can be characterized as solutions of a particular linear equation system, and show that, in general, a unique solution does not exist. In Subsection 3.4 we describe a special solution which minimizes the L2 norm (i.e., the Euklidean distance to the origin), which also leads us to a general solution, both obtained using the Moore-Penrose pseudoinverse. In Subsection 3.5 we consider approaches providing solutions that minimize the L1 norm (i.e., the sum of the absolute coefficients) and show that *shortestpath* and *randomwalk* belong to this class. In Subsection 3.6 we briefly refer to the 'L0 norm'. In Section 4 we describe a way to derive edge weights from path weights. In Section 5 we apply all approaches to the fictitious examples and to the real data set. Finally, the various approaches and their limitations are discussed (Section 6), an outlook is given, and recommendations for applying the methods in practice are provided in the conclusion (Section 7).

## 2 | REAL DATA EXAMPLE

We use a network meta-analysis of 22 treatments (including placebo and usual care) for primary care treatment of depression, see Figure 1.[12] The network consists of 93 studies, including 13 three-arm studies and one four-arm study. The primary outcome was 'Early response', measured as a binary variable. We use the odds ratio (OR) as effect measure. This data set is provided as Linde2016 in R package **netmeta**.[11]

## 3 | PATHS AND PATH WEIGHTS

Let $n$ be the number of treatments (nodes) in a network. Then $m = n(n-1)/2$ is the number of all possible comparisons of two treatments. Let $\mathbf{y}$ be a vector of length $m$ that contains the observed direct treatment effects, aggregated over all studies with the same comparisons, with zeros inserted where there are no direct comparisons.

### 3.1 | Hat matrix

The $m \times m$ full hat matrix $\mathbf{H}$ of the aggregate model maps $\mathbf{y}$ to the vector of estimated NMA effects $\hat{\theta}^{nma}$:

$$\hat{\theta}^{nma} = \mathbf{H}\mathbf{y} \qquad (1)$$

Here, $\mathbf{H}$ is defined as

$$\mathbf{H} = \mathbf{B}(\mathbf{B}^\top \mathbf{W} \mathbf{B})^+ \mathbf{B}^\top \mathbf{W}$$

where $\mathbf{B}$ is the edge-vertex incidence matrix of the full aggregate network (dimension $m \times n$), $\mathbf{W}$ is a diagonal matrix of inverse variance weights (dimension $m \times m$, setting weights for missing direct comparisons to 0), and $(\mathbf{B}^\top \mathbf{W} \mathbf{B})^+$ is the Moore-Penrose pseudoinverse of the Laplacian matrix $\mathbf{L} = \mathbf{B}^\top \mathbf{W} \mathbf{B}$.[10] We note that the elements of $\mathbf{W}$ can be assumed to be appropriately adjusted for multi-arm studies, as described in earlier publications.[13,14]

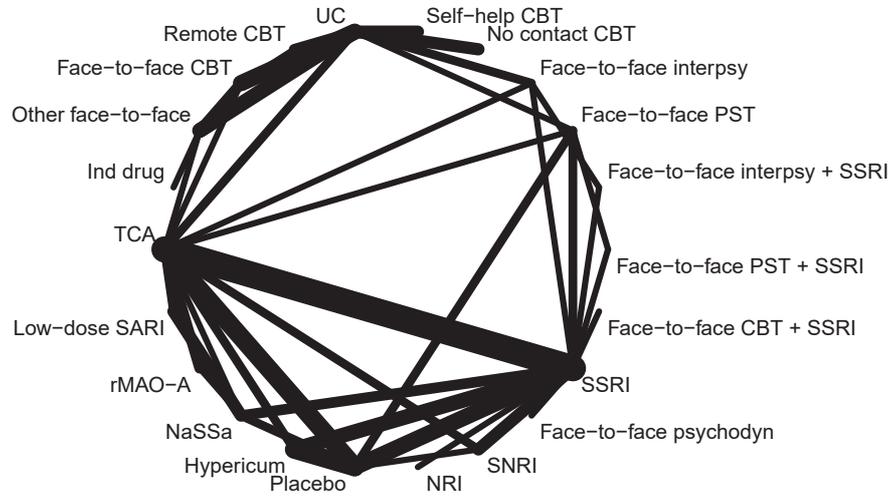

**Figure 1** Network graph for Linde 2016 data.

$\mathbf{H}$ is defined to contain zeros in columns that correspond to comparisons that are not part of the network. For simplicity of notation, we introduce the subscript $c$ for an arbitrary comparison, say A:B, comparing treatment A to treatment B. We may then write each row of (1) as

$$\hat{\theta}_c^{nma} = \mathbf{h}_c^\top \mathbf{y} \qquad (2)$$

where $\hat{\theta}_c^{nma}$ is the $c$'th entry of $\hat{\theta}^{nma}$ and $\mathbf{h}_c^\top$ denotes the $c$'th row of $\mathbf{H}$ (of length $m$). This means that $\hat{\theta}_c^{nma}$ is akin to a weighted sum of the direct treatment effect estimates $\mathbf{y}$, however, the 'weights' $h_{cj}$ ($j = 1, \ldots, m$) can be negative, and their sum is not always 1. These two properties interfere with the interpretation of the hat matrix elements as proportional contributions, and other ways to define contributions must be sought.

**Example 1**

We consider the network in Figure 2, left panel. All variances of direct comparisons are assumed to be 1. We have $m = 6$ comparisons, therefore the full hat matrix has dimension $6 \times 6$ and is given by

$$\mathbf{H} = \begin{pmatrix} AB & AC & AD & BC & BD & CD \\ 0.625 & 0.375 & 0 & -0.250 & -0.125 & 0.125 \\ 0.375 & 0.625 & 0 & 0.250 & 0.125 & -0.125 \\ 0.500 & 0.500 & 0 & 0.000 & 0.500 & 0.500 \\ -0.250 & 0.250 & 0 & 0.500 & 0.250 & -0.250 \\ -0.125 & 0.125 & 0 & 0.250 & 0.625 & 0.375 \\ 0.125 & -0.125 & 0 & -0.250 & 0.375 & 0.625 \end{pmatrix} \begin{matrix} A:B \\ A:C \\ A:D \\ B:C \\ B:D \\ C:D \end{matrix}$$

where the rows $c$ correspond to all possible comparisons and the columns to all possible edges (i.e., direct comparisons); the null column for edge AD signals that this edge does not exist in the network.

The sign of each coefficient in a row of **H** can be used to define the direction of the corresponding edge (which is initially arbitrary). Concentrating on a particular comparison $c = $ A:B, it is possible to change the direction of each 'negative' edge to ensure all coefficients to be non-negative. This induces a directed 'flow' from A (the 'source' of the path) to B (the 'sink').

To explain this, we concentrate on the first row of the hat matrix, that is comparison A:B. We find three positive (edges AB, AC, CD) and two negative entries (BC and BD). Accounting for the signs, we now interpret all edges as directed: $A \to B$, $A \to C$, $C \to D$, $C \to B$ and $D \to B$. The corresponding directed graph is seen in the right panel of Figure 1.

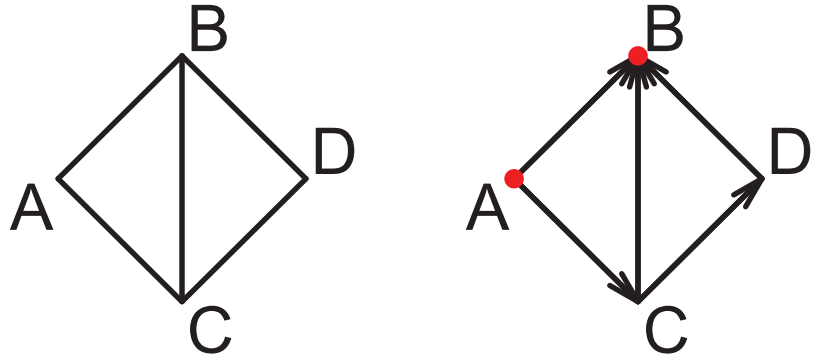

**Figure 2** Example 1: Fictitious example of 4 treatments A, B, C, D. Left panel: Network graph. Right panel: The same network as a directed graph for comparison A:B. The direction of each edge corresponds to the sign of its coefficients in $\mathbf{H}_{A:B}$.

## 3.2 | Path-design matrix

The two existing algorithms, *shortestpath* and *randomwalk*, both use the directed network defined by the hat matrix to define proportion contributions of each comparison in a network to each network estimate (what we call edge weights). They seek to find directed paths from A to B, adhering to the direction of edges as given by the hat matrix, and *randomwalk* will find *all P* paths from A to B. All these paths can be described by a $P \times m$ path-design matrix $\mathbf{Z}_c = \mathbf{Z}_{AB}$, where the rows correspond to all possible paths $p = 1, \ldots, P$ from A to B and the columns correspond to all possible (also indirect) comparisons in the network graph. The entries of the path-design matrix $\mathbf{Z}_{AB}$ are 1, −1, or 0, where 1 in row $p$ ($p = 1, \ldots, P$) and column $e$ means that path $p$ from A to B contains the directed edge $e$, −1 in row $p$ and column $e$ means that the directed path $p$ from A to B contains the reversely directed edge $e$, and 0 in row $p$ and column $e$ means that either there is no edge corresponding to comparison $e$, or edge $e$ is not on this path (in either direction).

Coming back to Example 1, the three (i.e., $P = 3$) possible paths from A to B are described by the path-design matrix

$$\mathbf{Z}_{AB} = \begin{pmatrix} AB & AC & AD & BC & BD & CD \\ 1 & 0 & 0 & 0 & 0 & 0 \\ 0 & 1 & 0 & -1 & 0 & 0 \\ 0 & 1 & 0 & 0 & -1 & 1 \end{pmatrix} \begin{array}{l} (A \to B) \\ (A \to C \to B) \\ (A \to C \to D \to B) \end{array} \quad (3)$$

**Example 2**

The second example with $n = 5$ treatments comes from Papakonstantinou and others and is shown in Figure 3 .[4, Suppl. 3] Again, all variances are assumed to be 1. Having $m = 10$ possible comparisons, we obtain the full $10 \times 10$ hat matrix $\mathbf{H}$ for the common effects model:

$$\mathbf{H} = \begin{pmatrix} AB & AC & AD & AE & BC & BD & BE & CD & CE & DE \\ 0.619 & 0 & 0 & 0.381 & -0.048 & -0.095 & -0.238 & -0.048 & 0 & -0.143 \\ \mathbf{0.571} & \mathbf{0} & \mathbf{0} & \mathbf{0.429} & \mathbf{0.571} & \mathbf{0.143} & \mathbf{-0.143} & \mathbf{-0.429} & \mathbf{0} & \mathbf{-0.286} \\ 0.524 & 0 & 0 & 0.476 & 0.190 & 0.381 & -0.048 & 0.190 & 0 & -0.429 \\ 0.381 & 0 & 0 & 0.619 & 0.048 & 0.095 & 0.238 & 0.048 & 0 & 0.143 \\ -0.048 & 0 & 0 & 0.048 & 0.619 & 0.238 & 0.095 & -0.381 & 0 & -0.143 \\ -0.095 & 0 & 0 & 0.095 & 0.238 & 0.476 & 0.190 & 0.238 & 0 & -0.286 \\ -0.238 & 0 & 0 & 0.238 & 0.095 & 0.190 & 0.476 & 0.095 & 0 & 0.286 \\ -0.048 & 0 & 0 & 0.048 & -0.381 & 0.238 & 0.095 & 0.619 & 0 & -0.143 \\ -0.190 & 0 & 0 & 0.190 & -0.524 & -0.048 & 0.381 & 0.476 & 0 & 0.429 \\ -0.143 & 0 & 0 & 0.143 & -0.143 & -0.286 & 0.286 & -0.143 & 0 & 0.571 \end{pmatrix} \begin{matrix} A:B \\ A:C \\ A:D \\ A:E \\ B:C \\ B:D \\ B:E \\ C:D \\ C:E \\ D:E \end{matrix}$$

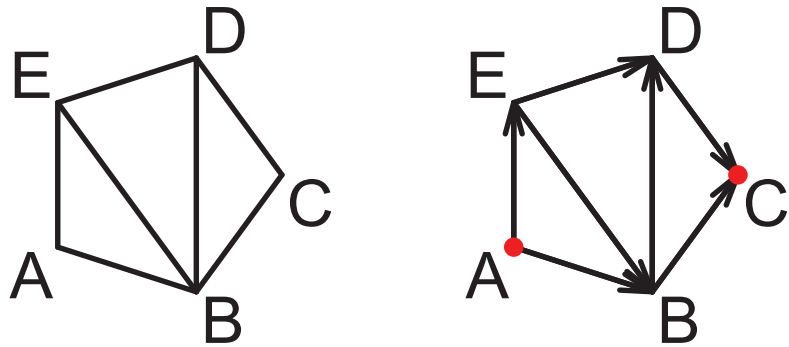

**Figure 3** Example 2: Fictitious example of 5 treatments A, B, C, D, E. Left panel: Network graph. Right panel: The same network as a directed graph for comparison A:C. The direction of each edge corresponds to the sign of its coefficients in $\mathbf{H}_{A:C}$.

We now focus on comparison A:C (printed in bold in $\mathbf{H}$) and interpret the signs in the corresponding row of $\mathbf{H}$ as directions. Adhering to these directions, paths from A to C are restricted to go through edges AB, AE, BC and BD in forward direction, whereas edges BE, CD and DE must be taken in backward direction (see Figure 2, right panel). Thus we obtain $P = 5$ directed paths from A to C, see Figure 4 .

The path-design matrix for comparison A:C is therefore:

$$\mathbf{Z}_{AC} = \begin{pmatrix} AB & AC & AD & AE & BC & BD & BE & CD & CE & DE \\ 1 & 0 & 0 & 0 & 1 & 0 & 0 & 0 & 0 & 0 \\ 1 & 0 & 0 & 0 & 0 & 1 & 0 & -1 & 0 & 0 \\ 0 & 0 & 0 & 1 & 1 & 0 & -1 & 0 & 0 & 0 \\ 0 & 0 & 0 & 1 & 0 & 1 & -1 & -1 & 0 & 0 \\ 0 & 0 & 0 & 1 & 0 & 0 & 0 & -1 & 0 & -1 \end{pmatrix} \begin{matrix} (A \to B \to C) \\ (A \to B \to D \to C) \\ (A \to E \to B \to C) \\ (A \to E \to B \to D \to C) \\ (A \to E \to D \to C) \end{matrix}$$

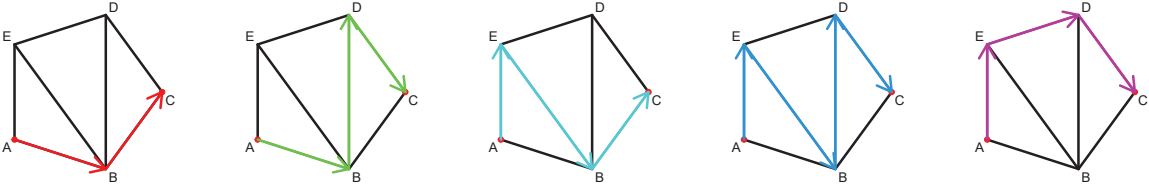

**Figure 4** Example 2: All directed paths from A to C with direction compatible with the signs in $H_{AC}$.

## 3.3 | Finding path weights

König et al. interpreted the rows of the hat matrix as a directed 'evidence flow'.[7] Papakonstantinou et al. extended this concept and suggested to decompose the evidence flow on the directed network as shown in the examples above, say from A to B, into parts called streams on the various paths from A to B, and defined edge weights based on this.[4] Davies et al. adopted this idea when defining edge weights based on random walks.[10] Generalizing these thoughts, we aim to construct weights for each path that contributes to the flow of evidence from A to B. This can be formalized in the following way. Given a comparison $c$, we are looking for a vector of path weights $\phi_c$ of length $P$ such that path $p$ obtains weight $\phi_{cp}$. We can rewrite equation (2) as

$$\hat{\theta}_c^{nma} = \mathbf{h}_c^\top \mathbf{y} = \phi_c^\top \mathbf{Z}_c \mathbf{y}$$

for all $\mathbf{y}$. The right equation holds if $\phi_c$ is a solution of the linear equation system

$$\mathbf{h}_c^\top = \phi_c^\top \mathbf{Z}_c. \tag{4}$$

In general, $\mathbf{Z}_c$ has no inverse and thus there is not always a unique solution. Following a theorem given by Albert[15, (3.12)], this equation has a solution if and only if

$$\mathbf{h}_c^\top \mathbf{Z}_c^+ \mathbf{Z}_c = \mathbf{h}_c^\top$$

where $\mathbf{Z}_c^+$ is the Moore-Penrose pseudoinverse of $\mathbf{Z}_c$. $\mathbf{Z}_c^+$ is a $m \times P$ matrix. In fact, we already know that the algorithms *shortestpath* and *randomwalk* provide possible exact solutions. Following a theorem given by Albert, the general solution space of the equation system (4) can be written as[15, Theorem (3.12)]

$$\phi_c^\top = \mathbf{h}_c^\top \mathbf{Z}_c^+ + \mathbf{x}^\top (\mathbf{I} - \mathbf{Z}_c \mathbf{Z}_c^+) \tag{5}$$

where $\mathbf{I}$ is the $P \times P$ identity matrix and $\mathbf{x} \in \mathbb{R}^P$ an arbitrary vector of length $P$. The solution is unique if and only if $\mathbf{Z}_c \mathbf{Z}_c^+ = \mathbf{I}$.[15, (3.12)(c)]

**Proposition 1.** The sum of entries in each solution $\phi_c$ of (4), written as in (5), is 1.

That the coefficients of $\phi_c$ sum up to 1 does not mean that all entries are non-negative. They can be (and often are) negative, as we will see for Example 2. We refer to Appendix A.1 for a proof of the proposition and to Appendix A.2 for further remarks.

## 3.4 | A solution minimizing the L2 norm

Based on the general solution (5) of the linear equation (4) described in the previous subsection, we may look at special cases. One of them is obtained by setting $\mathbf{x} = 0$ in 5:

$$\phi_c^\top = \mathbf{h}_c^\top \mathbf{Z}_c^+. \tag{6}$$

This solution has the special property that it minimizes the Euklidean (L2) distance from the origin (i.e., the sum of the squared coefficients, $\phi_c^2$).[15, Theorem (3.4)] We refer to this approach and its solution as *pseudoinverse*.

## 3.5 | Solutions minimizing the L1 norm

We may also investigate an alternative criterion, which is to minimize the *sum* of the absolute coefficients $|\phi_c|$, which means minimizing with respect to the L1 norm.

We already know that the set of all solutions for (4) is given by (5). Thus, we are now looking for $\mathbf{x} \in \mathbb{R}^p$ to be inserted in (5) such that $\sum |\phi_{ci}|$ is minimal under condition (4). Technically, an L1 solution for this problem can be found using function l1() from R package **cccp** for convex programming,[16] setting $\mathbf{A}_c = \mathbf{I} - \mathbf{Z}_c \mathbf{Z}_c^+$ and minimizing $||\phi_c^\top + \mathbf{x}^\top \mathbf{A}_c||_1$ where $\phi_c$ is the *pseudoinverse* solution given in (6). The coefficients of an L1 solution $\phi$ cannot be negative; see Appendix A.3 for a proof. We refer to this approach as *cccp*.

Also *shortestpath* and *randomwalk* provide L1 solutions. This is because they have exclusively non-negative coefficients $\phi$ by construction.[4,10] Hence the sum and the sum of the absolute values agree, and both are 1 (as for all solutions), which is the minimal possible value. It also follows that, in general, the L1 solution is not unique. The set of L1 solutions is a subset of the affine set described in (5) characterized by all coefficients of $\phi_c$ being non-negative.

The solutions need not be equal to each other and do not always lead to the same weights, as shown by Davies et al. (2022) for *shortestpath* and *randomwalk*.[10] Thus, obviously, though an exact solution of (4) does always exist, it is in general not unique. Indeed, we will see that all four solutions, *pseudoinverse*, *shortestpath*, *randomwalk*, and *cccp* can be different.

We note that all solutions are exact and thus trivially minimize the Euklidean distance L2 between $\mathbf{h}_c^\top$ and $\phi_c^\top \mathbf{Z}_c$: the distance is simply 0. In general the linear system (4) is under-determined. The *pseudoinverse* solution (6) has the additional property that $\phi_c$ minimizes the Euklidean distance L2 to the origin in $\mathbb{R}^P$.

## 3.6 | Solutions minimizing the 'L0 norm'

We also searched for a solution of (4) that minimizes the so-called 'L0 norm' (which is not a norm in the mathematical sense), here defined as the number of non-zero coefficients in $\phi_c$.[17] We suspected that this may hold for *shortestpath*, but found a counterexample where a different order of selecting paths led to a smaller number of non-zero coefficients. See Appendix A.4 for Example 2 and the counterexample and Appendix A.5 where we give an upper limit for the number of paths needed by approaches such as *shortestpath*.

## 4 | EDGES AND EDGE WEIGHTS

The original motivation for defining path weights was the need to define the proportion contribution of any edge in the network to each network estimate. However, it is not immediately clear how to proceed from path weights to edge weights, or what other sensible ways of defining edge weights are possible that are not based on path weights.

## 4.1 | Path-based edge weights - equal shares

Papakonstantinou et al. suggested to use the path weights $\phi_c$ from the *shortestpath* approach to obtain the contributions of any edge in the network to a given comparison $c$.[4] Their idea was to distribute the weight of each path in equal shares to all edges that constitute that path and then, for each edge, sum over all paths that contain that edge. Formally, the entries of $\phi_c$ are transformed to edge weights, forming a vector $\mathbf{w}_c$ of length $m$, by setting

$$\mathbf{w}_c^\top = (\phi_c / \mathbf{l}_c)^\top |\mathbf{Z}_c| \tag{7}$$

where $|\mathbf{Z}_c|$ is the matrix derived from $\mathbf{Z}_c$ by taking elementwise absolute values, $\mathbf{l}_c$ is the vector of path lengths, given by the row sums of $|\mathbf{Z}_c|$, and the division $\phi_c / \mathbf{l}_c$ is also made elementwise. The sum of these weights (7) is always 1 due to

$$\mathbf{w}_c^\top \mathbf{1} = (\phi_c / \mathbf{l}_c)^\top |\mathbf{Z}_c| \mathbf{1} = (\phi_c / \mathbf{l}_c)^\top \mathbf{l}_c = \phi_c^\top \mathbf{1} = 1$$

(where vector $\mathbf{1}$ is the vector of ones of length $m$).

This method can be applied to any solution $\phi$ of equation (4), leading to potentially different weights. For example, Davies et al. used the same method to define weights based on the *randomwalk* approach.[10] For both the *shortestpath* and the *randomwalk* approach all weights are non-negative. For the *pseudoinverse* approach, we sometimes observed (small) negative weights. This phenomenon will be further discussed below.

**Path weights from edge weights**

The edge weights can be derived from the path weights $\phi_c$ using (7). However, it is in general not possible to reverse this process. Edge weights do not necessarily determine path weights. Indeed, different sets of path weights may lead to the same edge weights, and a given set of edge weights might even correspond to no set of path weights at all.

## 5 | RESULTS

To apply all methods, we added *cccp* and *pseudoinverse* as new values of argument `method` to function `netcontrib` of R package **netmeta**. An R script and data to reproduce all results is found as Supporting Information 8. First, we give the results for the two fictitious examples.

### 5.1 | Example 1

All path-based methods result in the same path weights $\phi_{AB}$ for the three paths:

$$\phi_{AB} = (0.625, 0.25, 0.125) = (5, 2, 1)/8,$$

providing the edge weights $(0.625, 0.16667, 0.125, 0.041667, 0.041667) = (15, 4, 3, 1, 1)/24$ for edges AB, AC, BC, BD, CD (note that there is no edge AD and therefore comparison A:D contributes 0). Also for all other comparisons the weights, shown in Table 1, agree between methods.

**Table 1** Edge weights of all path-based approaches for Example 1.

|     | A:B   | A:C   | B:C   | B:D   | C:D   |
|-----|-------|-------|-------|-------|-------|
| A:B | **0.625** | 0.167 | 0.125 | 0.042 | 0.042 |
| A:C | 0.167 | **0.625** | 0.125 | 0.042 | 0.042 |
| A:D | 0.250 | 0.250 | 0.000 | 0.250 | 0.250 |
| B:C | 0.125 | 0.125 | **0.500** | 0.125 | 0.125 |
| B:D | 0.042 | 0.042 | 0.125 | **0.625** | 0.167 |
| C:D | 0.042 | 0.042 | 0.125 | 0.167 | **0.625** |

### 5.2 | Example 2

For this example, the path-based methods result in different path weights (Table 2) and also lead to different edge weights (Table 3). As expected, the shortest path ABC obtains the largest weight for all methods. The *shortestpath* approach puts relatively large weight on the longest path AEBDC, whilst ignoring paths ABDC and AEBC. This is because exhausting the flow of the ABC path means removing edges AB and BC that would be needed for ABDC or AEBC. The flow in the edges of those paths is then assigned to the remaining paths. By contrast, for *pseudoinverse* the coefficient that corresponds to the longest path AEBDC is negative. We also used a variant of the *shortestpath* algorithm ('other path order') where we did not start with the shortest path (ABC), but with path AEDC, followed by AEBC, ABDC, and ABC, thus avoiding the longest path AEBDC. This again led to a different result. Due to symmetry, the weights for ABDC and AEBC agree within each method. All methods provided the same weight for path AEDC.

Although the path weights (Table 2) differ markedly between methods, the resulting edge weights (Table 3) are similar.

The path-design matrix $\mathbf{Z}_{AC}$ ($P = 5$ paths) has rank 4 and thus the space of solutions of (4) has rank $5 - 4 = 1$, that is, all solutions lie on one line in $\mathbb{R}^5$. For this example it can be shown that *shortestpath* is the unique solution that minimizes the 'L0 norm'. We give a proof in Appendix A.4 .

**Table 2** Path weights for comparison A:C for Example 2, different methods. Shortest path order: ABC, AEDC, AEBDC. Other path order: AEDC, AEBC, ABDC, ABC.

| Method | Path | | | | |
|---|---|---|---|---|---|
| | ABC | ABDC | AEBC | AEBDC | AEDC |
| *shortestpath* | 0.571 | 0 | 0 | 0.143 | 0.286 |
| Other path order | 0.429 | 0.143 | 0.143 | 0 | 0.286 |
| *randomwalk* | 0.457 | 0.114 | 0.114 | 0.029 | 0.286 |
| *cccp* | 0.471 | 0.101 | 0.101 | 0.042 | 0.286 |
| *pseudoinverse* | 0.393 | 0.179 | 0.179 | -0.036 | 0.286 |

**Table 3** Path-based edge weights for comparison A:C for Example 2, different methods. Shortest path order: ABC, AEDC, AEBDC. Other path order: AEDC, AEBC, ABDC, ABC.

| Method | Comparison | | | | | | |
|---|---|---|---|---|---|---|---|
| | A:B | A:E | B:C | B:D | B:E | C:D | D:E |
| *shortestpath* | 0.286 | 0.131 | 0.286 | 0.036 | 0.036 | 0.131 | 0.095 |
| Other path order | 0.262 | 0.143 | 0.262 | 0.048 | 0.048 | 0.143 | 0.095 |
| *randomwalk* | 0.267 | 0.140 | 0.267 | 0.045 | 0.045 | 0.140 | 0.095 |
| *cccp* | 0.269 | 0.139 | 0.269 | 0.044 | 0.044 | 0.139 | 0.095 |
| *pseudoinverse* | 0.256 | 0.146 | 0.256 | 0.051 | 0.051 | 0.146 | 0.095 |

## 5.3 | Real data example

We applied *shortestpath*, *randomwalk*, *pseudoinverse* and the L1 method as implemented in R package **cccp** to calculate the edge weights for the depression data.[12] R code and results are given as Supporting information 8. We note that function `netcontrib()` does not provide the path weights for *shortestpath* and *randomwalk*, only the resulting edge weights. As there are $n = 22$ interventions, for each method the edge weights are summarized in a matrix with $m = n(n-1)/2 = 231$ rows (one for each pair of different nodes in the network) and 41 columns (one for each pairwise comparison for the aggregate network, i.e., studies with the same comparison counted only once). It was confirmed that both the sum of the path weights $\phi$ and the sum of the edge weights **w** was 1 for all methods. For all methods, the same set of 1610 of $231 * 41 = 9471$ (17 %) edge weights were zero for all methods. An additional 50 edge weights (0.53 %) obtained negative values near zero for the *pseudoinverse* method; the minimum value was -0.0002128679. It seems improbable that this was due to a numerical problem, as the sum of all weights was exactly 1 for each comparison.

As expected, we empirically observed a negative correlation between path length and coefficients of $\phi$ for *pseudoinverse*: coefficients become negative for 'less important' long paths, as indicated by Figure 5. This is in line with earlier empirical observations.[6]

Figure 6 shows Bland-Altman plots of all $231 * 41 = 9471$ edge weights for each pair of methods. We see large differences, particularly for 'small' (in the sense of the average over both compared methods) contribution values (i.e., up to about 0.4), between the *pseudoinverse* method on the one hand and *shortestpath*, *randomwalk* and *cccp* on the other hand (upper panels). The *pseudoinverse* method tends to overemphasize very small weights and to downweight moderate contributions compared to the other methods, where the differences were maximal for contribution values near 0.3. The differences between *shortestpath* and *randomwalk* (or *cccp*) were smaller, but still present for small contribution values. The *cccp* and the *randomwalk* approach showed the greatest similarity (right bottom panel). We will revisit these observed differences in the Discussion.

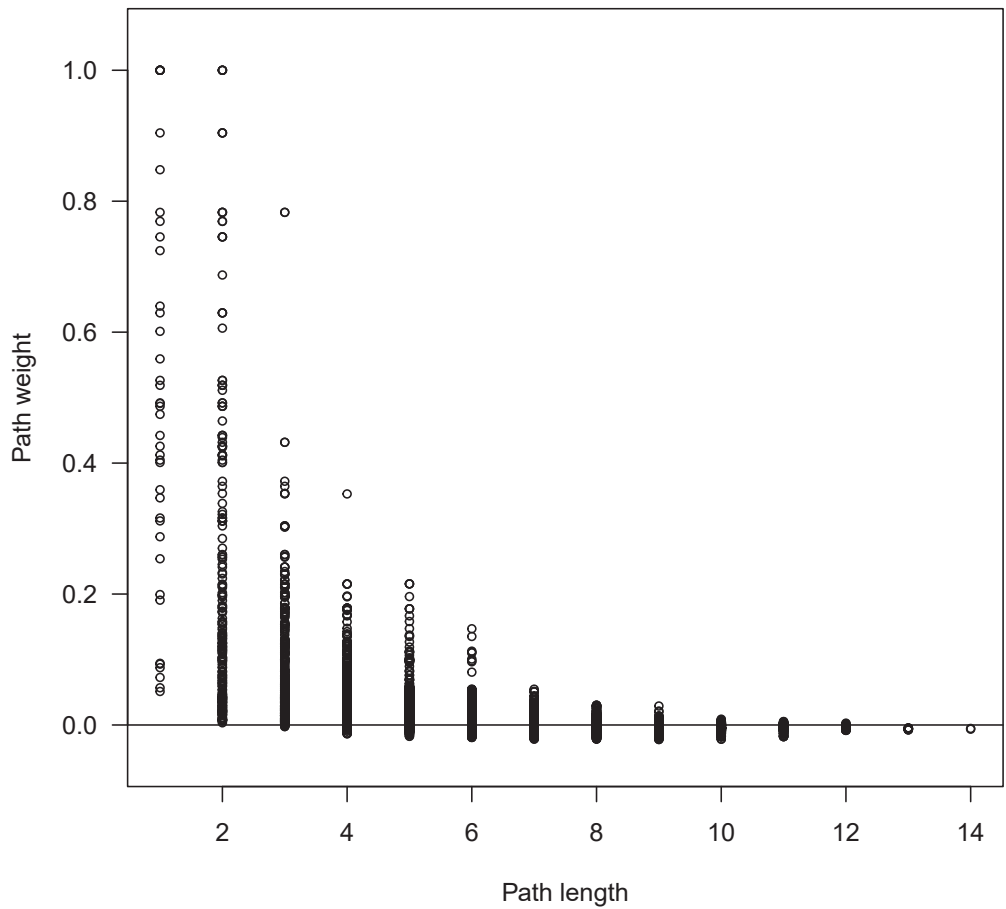

**Figure 5** Path weights against path lengths for *pseudoinverse* approach (Linde2016 data). For any of the 231 possible comparisons, all paths are considered and each dot corresponds to one path. There are 26490 paths in total. Note the occurrence of negative path weights.

## 6 | DISCUSSION

Various suggestions have been made to measure the proportional (or percentage) contributions of each comparison (or, alternatively, each study) to each NMA estimate, all based on different versions of the hat matrix[1,4,10]. It turned out that different approaches led to different results except in very simple networks, though the differences were mostly not very large, particularly for very highly contributing edges[4,10]. In our real data example, we observed the largest differences for contributions around 30 %. This was similar for other real data examples (not shown).

In this article, we focussed on path-based approaches, derived from the hat matrix. We have shown that the path-based methods, as proposed so far, are special cases of a general concept, based on the path-design matrix, and that in general there is a space of infinitely many solutions. All considered solutions are exact. The approaches *shortestpath*, *randomwalk* and *cccp* turned out to be solutions with respect to the L1 norm, whereas the *pseudoinverse* approach minimizes the L2 norm. Because of these different criteria, it is not surprising that for our examples the largest difference is often seen between the *pseudoinverse* on the one hand and all others on the other hand. In general, the three described L1 solutions are only three of a potentially infinite subspace of L1 solutions. An open question is what, if any, are the special features of the solution *cccp* provided by the R function `l1()`. We did not find any description of these on the (extremely short) help page of `l1()`.

**Figure 6** Bland-Altman plots for comparing edge weights from different approaches (Linde2016 data). Each panel compares two approaches, showing the difference of all edge weights using one approach compared to another against the mean of both values. Top left: *pseudoinverse - shortestpath*, top middle: *pseudoinverse - randomwalk*, top right: *pseudoinverse - cccp*, bottom left: *randomwalk - shortestpath*, bottom middle: *cccp - shortestpath*, bottom right: *cccp - randomwalk*.

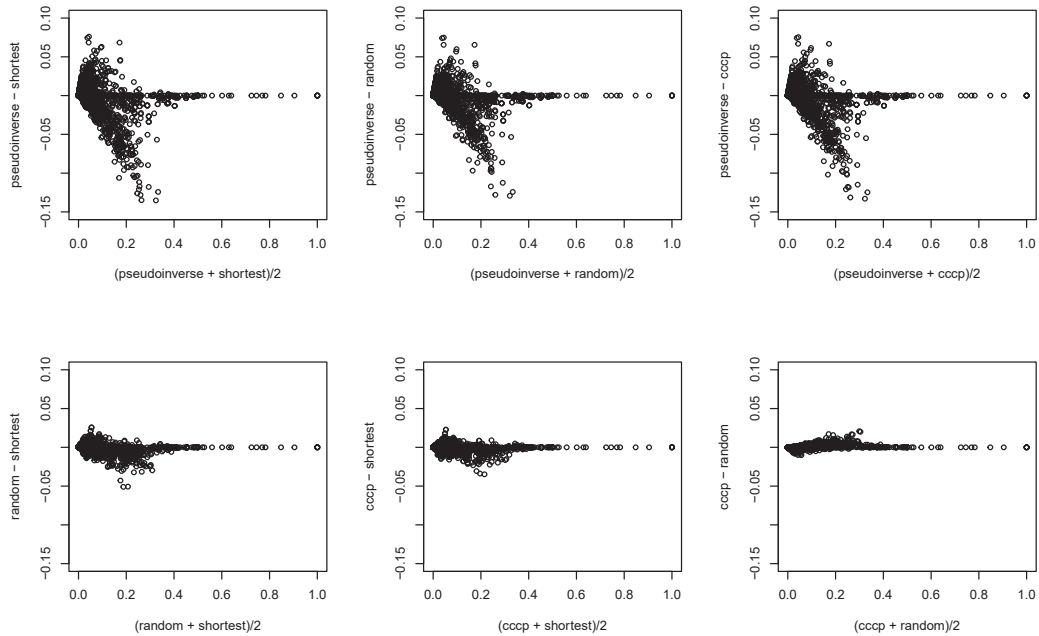

**Variance of the path weights**   The *pseudoinverse* approach minimizes the sum of squares of the path weights $\phi_c$, and because their sum is always 1, this means that it minimizes the variance of the path weights. This in turn often (but not always) leads to smaller variance between edge weights. In other words, the *pseudoinverse* approach tends to downweight the short paths and to upweight long paths/less important edges compared to the other approaches. By contrast, *shortestpath* often shows the largest variance between path weights. The reason is that this algorithm starts from the shortest paths, quickly exhausting their flow and thus tends to ignore longer paths for more complex networks, such that these obtain path weight zero. *shortestpath* and *pseudoinverse* differ most, *randomwalk* and *cccp* are most similar.

**Negative weights**   We observed negative values for $\phi$ for the *pseudoinverse* method, typically seen for long and less important paths. This method seems to penalize these paths. Nevertheless, the resulting edge weights **w** were mostly positive also for *pseudoinverse*, with a number of exceptions where we found negative weights very near to zero. Though this has not much practical relevance, it does not seem to be a merely numerical problem and thus makes these numbers difficult to interpret as contributions.

**Run time**   A requirement for the approaches *randomwalk*, *pseudoinverse*, and *cccp* is an all-paths search. This can be very time-consuming for large networks. For the Linde2016 data (26490 paths in total), *randomwalk* needed 375 sec, compared to 47 sec with *shortestpath* on a notebook under Windows operating system. For another network with 63 treatments on postoperative nausea and vomiting,[18] *randomwalk* and *pseudoinverse* needed over 5 days on a single core of a Xeon E5-2643 v3 CPU (results not shown). The fastest method is *shortestpath* (currently default in CINeMA and R package **netmeta**). This is the only path-based method that does not need an all-paths search; e.g., for our Example 2, *shortestpath* ignores paths $A \rightarrow B \rightarrow D \rightarrow C$ and $A \rightarrow E \rightarrow B \rightarrow C$. Ignoring paths means setting their $\phi$ coefficients to zero and thus reducing the 'L0 norm'. For the other three path-based approaches, the all-paths search is the bottleneck with respect to run time.

**Comparing the approaches**   To give a recommendation which method is to be preferred, we thought of the criteria interpretability, distinctness, and practicability (implementation and run time). The occurrence of negative edge weights, which are difficult to interpret, provides an argument against *pseudoinverse*. Another criterion is the ability of a method to distinguish important and less important contributions. A measure for this is the variance of the path weights, which should be large, and is typically largest for *shortestpath*. Long run times for larger networks argue against *randomwalk*, *pseudoinverse*, and *cccp*. At the moment, we do not see a way to speed up these three methods by avoiding an all-paths search. Accordingly, the *shortestpath* approach is the default in R function `netcontrib`. Table 4 gives an overview of the properties of the different approaches.

**Table 4** Comparison of approaches for finding path weights.

| Method | Minimizes norm | Variance | Negative weights | All-paths search needed |
| --- | --- | --- | --- | --- |
| *pseudoinverse* | L2 | smallest | yes | yes |
| *shortestpath* | L1 | largest | no | no |
| *randomwalk* | L1 | medium | no | yes |
| *cccp* | L1 | medium | no | yes |

**Edge weights** To proceed from path weights to edge weights, we divided each path contribution into equal shares for all its edges. We acknowledge that this idea is not compulsory, and in fact it is even controversial, see, for example, the openly available peer review by König to the first version of Papakonstantinou and others.[4] However, we do currently not see a convincing alternative to derive edge weights from path weights.

**Alternative approaches** Instead of defining edge contributions as percentages, it has been proposed to define the importance of each comparison or each study based on their variances.[5] These importances do not necessarily add to 1. For example, suppose a network structure that is a chain of treatments A, B, C where A is only compared to B and B is only compared to C. For the indirect comparison A:C this approach would assign an importance of 1 to both edges. The interpretation is that both comparisons are 100 % necessary to make the indirect comparison possible. Adding these importances to 2 does not make sense.

We did not discuss the 'borrowing of strength' approach which is based on a decomposition of the score statistic.[2,3]

# 7 | CONCLUSION

Originally motivated by the need to define edge weights, we developed a framework for defining path weights. Path-based contributions (edge weights) are in general not uniquely defined. We have presented a general framework, based on the path-design matrix, that includes two known approaches *shortestpath* and *randomwalk*. Further, we have shown that there are different criteria (L1 and L2 minimization of the path weights) and that the known approaches as well as *cccp* meet L1, while *pseudoinverse* meets L2. The *pseudoinverse* approach may lead to negative contributions, which makes it difficult to interpret. For complex networks, *shortestpath* has been shown to be superior with respect to run time, compared to the other approaches, and is thus recommended in practice.

The path-weights framework is promising also in answering other research questions in NMA, for example, to define inconsistency between paths, correlation between paths and other novel concepts based on paths. These concepts are currently being developed and have the potential to lead to new insights for NMA.

# ACKNOWLEDGMENTS

AN and TP were supported by the Emmy Noether Programme of the German Research Foundation (grant number NI 2226/1-1). ALD was funded by the Engineering and Physical Sciences Research Council (EPSRC UK), grant number EP/R513131/1.

**Author contributions**

ALD conceived the path-design matrix. GR did the mathematical derivations given in Section 3 and the Appendix, wrote the R functions for *pseudoinverse* and *cccp*, analyzed all data and wrote the first draft of the paper. ALD, AN, GS, TG and TP critically contributed to the manuscript. GR and TP drafted the proof in A.5. GS implemented *pseudoinverse* and *cccp* in R package **netmeta** and contributed to the analyses. All authors read and approved the final manuscript

**Financial disclosure**

None reported.

**Conflict of interest**

The authors declare no potential conflict of interests.

# 8 | SUPPORTING INFORMATION

The following supporting information is available as part of the online article:
**S1: R-Code-Submission.R** R code for running all analyses and producing all figures.
**S2: Linde2016-shortestpath-result.rda** R data file with results for real data example.
**S3: Linde2016-randomwalks-result.rda** R data file with results for real data example.
**S4: Linde2016-L1result.rda** R data file with results for real data example.
**S5: Linde2016-L2result.rda** R data file with results for real data example.

> **How to cite this article:** Rücker G., T. Papakonstantinou, A. Nikolakopoulou, G. Schwarzer, T. Galla, and A. L. Davies (2023), Shortest path or random walks? A framework for path weights in network meta-analysis, , *2023;00:1–16*.

# APPENDIX

# A PROOFS

## A.1 Proof of the proposition

*Proof.* Let **1** be the vector of ones of length $P$. We want to show that

$$\phi_c^\top \mathbf{1} = \mathbf{h}_c^\top \mathbf{Z}_c^+ \mathbf{1} = 1$$

for each solution $\phi_c$ in equation (5).

**Step 1**

We first note that it is sufficient to show that

$$\mathbf{Z}_c \mathbf{Z}_c^+ \mathbf{1} = \mathbf{1} \quad \text{and thus} \quad (\mathbf{I} - \mathbf{Z}_c \mathbf{Z}_c^+) \mathbf{1} = \mathbf{0}.$$

This is because if this holds, the second term in (5) cancels when applied to **1**, giving

$$\phi_c^\top \mathbf{1} = \mathbf{h}_c^\top \mathbf{Z}_c^+ \mathbf{1},$$

independently of the choice of **x**. In other words, all solutions for $\phi_c$ have the same sum $\phi_c^\top \mathbf{1}$ over their entries. As we know the sum to be 1 for both *shortestpath* and *randomwalk*, it must be 1 for any other solution and particularly for the *pseudoinverse* solution, see the original papers by Papakonstantinou[4, page 5, "Streams"] and Davies[10, 5.1].

**Step 2**
From general rules for the pseudoinverse we know:[15]

1. $\mathbf{Z}_c \mathbf{Z}_c^+$ is idempotent because $\mathbf{Z}_c \mathbf{Z}_c^+ \mathbf{Z}_c \mathbf{Z}_c^+ = \mathbf{Z}_c \mathbf{Z}_c^+$.

2. Thus, $\mathbf{Z}_c \mathbf{Z}_c^+$ is a projection matrix [15, (3.5), (3.7.6)].

3. For all vectors $\mathbf{x}$, $\mathbf{Z}_c \mathbf{Z}_c^+ \mathbf{x}$ is the projection of $\mathbf{x}$ on the range of $\mathbf{Z}_c$.[15, (3.5)]

4. Because $\mathbf{Z}_c \mathbf{Z}_c^+$ is a projection matrix, all its eigenvalues $\lambda_i$ are either 1 or 0.[15, (3.7.4)]

5. Due to the properties of the pseudoinverse, $\mathbf{Z}_c \mathbf{Z}_c^+$ is also symmetric and therefore normal, i.e., there exists an orthonormal basis of $\mathbb{R}$ of eigenvectors $\mathbf{x}_1, \ldots, \mathbf{x}_P$ of $\mathbf{Z}_c \mathbf{Z}_c^+$.[15, (3.9.1)]

6. The vector of ones, $\mathbf{1}$, can be written as
$$\mathbf{1} = \sum_{i=1}^{P} (\mathbf{1}^\top \mathbf{x}_i) \mathbf{x}_i = \sum_{i=1}^{P} \sum_{j=1}^{P} x_{ij} \mathbf{x}_i$$
where $\mathbf{x}_1, \ldots, \mathbf{x}_P$ are the eigenvectors of $\mathbf{Z}_c \mathbf{Z}_c^+$ with coefficients $x_{ij}$.

7. We then have
$$\mathbf{Z}_c \mathbf{Z}_c^+ \mathbf{1} = \sum_{i=1}^{P} \sum_{j=1}^{P} x_{ij} \mathbf{Z}_c \mathbf{Z}_c^+ \mathbf{x}_i = \sum_{i:\lambda_i=1} \sum_{j} x_{ij} \mathbf{x}_i$$
Because of item 3 above, $\mathbf{Z}_c \mathbf{Z}_c^+ \mathbf{1}$ is the projection of $\mathbf{1}$ on the range of $\mathbf{Z}_c$.

Consider the set $M$ of edges starting in node A.[1] As each path from A to B begins in exactly one of these edges, each row of $\mathbf{Z}_c$, corresponding to a path, has entry 1 in exactly one of the columns belonging to $M$. Now define a vector $\mathbf{y}$ of length $m$ with ones in the positions belonging to the edge set $M$ and zeros otherwise. By construction, matrix $\mathbf{Z}_c$ maps $\mathbf{y}$ to the sum of matrix elements over just the columns belonging to $M$. As there is exactly one 1 in each of these columns and zeros otherwise, this sum is always one and we have
$$\mathbf{Z}_c \mathbf{y} = \mathbf{1}.$$
This means that $\mathbf{1}$ is in the range of $\mathbf{Z}_c$ and thus it means equality of the expressions in items 6 and 7 above: the row sums of $\mathbf{Z}_c \mathbf{Z}_c^+$ are 1:
$$\mathbf{Z}_c \mathbf{Z}_c^+ \mathbf{1} = \mathbf{1} \quad \text{and} \quad (\mathbf{I} - \mathbf{Z}_c \mathbf{Z}_c^+) \mathbf{1} = \mathbf{0}.$$
This is what we wanted to prove.[2] □

## A.2 Some further remarks

*Remark 1.* Equation (4) leads to an interesting interpretation of the sum of the absolute elements in a row $\mathbf{h}_c$ of the hat matrix. Without loss of generality, we may assume that all elements of row $c$ are non-negative (otherwise we can change the orientation of all edges with negative entry). Let $\mathbf{l}_c$ be the vector of path lengths, given by the row sums of $\mathbf{Z}_c = |\mathbf{Z}_c|$. Then the row sum is
$$\mathbf{h}_c^\top \mathbf{1} = \boldsymbol{\phi}_c^\top \mathbf{Z}_c \mathbf{1} = \boldsymbol{\phi}_c^\top \mathbf{l}_c = \sum_{p=1}^{P} \phi_{cp} l_{cp}$$
which can be interpreted as the average path length for comparison $c$, weighted by the path weights $\boldsymbol{\phi}_c$. This makes it plausible and even necessary that most of these row sums are greater than 1. The average path length does not depend on $\boldsymbol{\phi}_c$.

*Remark 2.* The space consisting of all vectors whose coefficients sum up to 1 is a $(P-1)$-dimensional affine subspace of $\mathbb{R}^P$ (i.e., a hyperplane), and the subspace of solutions of (4) is a subspace of this hyperplane with dimension $P - \text{rank}(\mathbf{Z}_c \mathbf{Z}_c^+)$.

*Remark 3.* It must be noted that some formal solutions of (5) may not make sense as path weights. For example, weights may be counterintuitively small or large, and the equation does not account for any symmetry in the network (though rarely occurring in practice). We discuss special solutions in Sections 3.4 - 3.6 of the main text.

---

[1]The same argument works if we consider all edges ending in node B. What we need is a minimum cut in the sense of the theorem by Ford and Fulkerson.[19]
[2]Because $\mathbf{Z}_c \mathbf{Z}_c^+$ is symmetric (item 5 above), the column sums are also 1, such that $\mathbf{Z}_c \mathbf{Z}_c^+$ is a double-stochastic matrix.

## A.3 Proof that the coefficients $\phi$ of a L1 solution cannot be negative

Assume there were any negative coefficients in $\phi$. We may write $\phi_+$ for the sum of non-negative coefficients and $-\phi_-$ for the sum of negative coefficients. As the sum is 1, we have $\phi_+ - \phi_- = 1$ and thus $\phi_+ = 1 + \phi_- > 1$. The sum of the absolute coefficients is then $\phi_+ + \phi_- = 1 + 2\phi_- > 1$. This cannot be a solution with respect to the L1 norm, because it is beaten by *shortestpath* and *randomwalk* (both have non-negative coefficients that sum up to 1).

## A.4 Proof that *shortestpath* minimizes the 'L0 norm' for Example 2 , but not in general

**Example 2**

For Example 2 we show that *shortestpath* minimizes the number of non-zero coefficients of $\phi_{AC}$, sometimes called the 'L0 norm'. We know that the space of solutions for this example has dimension 1. Starting from two solutions, for example *shortestpath* ($\phi_{shortest}$) and *randomwalk* ($\phi_{random}$) , we can thus write the general solution

$$\phi = \phi_{shortest} + \lambda(\phi_{random} - \phi_{shortest})$$

with $\lambda \in \mathbb{R}$. Coefficient $i$ of $\phi$ becomes zero if

$$\lambda = \phi_{shortest,i}/(\phi_{shortest,i} - \phi_{random,i})$$

Using $\phi_{shortest} = (0.571, 0, 0, 0.143, 0.286)$ and $\phi_{random} = (0.457, 0.114, 0.114, 0.029, 0.286)$ (see Table 2) we obtain

$$\phi_{shortest}/(\phi_{shortest} - \phi_{random}) = (5, 0, 0, 1.25, \infty)$$

This means that setting $\lambda$ to one of these values provides a zero coefficient at the corresponding position in $\phi$ (it does not make sense to set $\lambda$ to $\infty$, though). Setting $\lambda = 5$ gives zero weight to path ABC, which is not a very intuitive solution. Setting $\lambda = 0$ provides the *shortestpath* solution with two zeros in the second (ABDC) and third (AEBC) position. Setting $\lambda = 1.25$ provides a zero in the fourth position (AEBDC) and corresponds to the 'other path' solution. These are the only solutions with zero path weights.

**A counterexample**

We consider a network with the same structure as Example 2, but different variances. Setting the variances of the edges AB, AE, BC, BD, BE, CD, ED to 4, 3, 8, 1, 2, 2, 4 (in this order), the hat matrix row for comparison A:C becomes as shown in Figure A1 , interpreted as a total flow of 1 from A to C. The *shortestpath* approach starts with the shortest path from A to C, which is ABC and obtains weight 0.25. The next paths are ABDC and AEDC, both with weight 0.25. The flow is exhausted with the longest path AEBDC which also gets weight 0.25. Path AEBC is not needed and gets weight 0.

Alternatively, we could start with the 'widest road', which is path ABDC (weight 0.5), continuing with AEBC and AEDC (both weight 0.25) and thus exhausting the flow using only three paths, avoiding the shortest path ABC and the longest path AEBDC. Thus it is not *shortestpath* that minimizes the number of non-zero coefficients (the 'L0 norm'). Also the variance of the path weights is larger for this approach (0.04375) than for *shortestpath* (0.0125). The average path length $\sum h_{A:C,i}$ is 3 for all solutions, see A.2 .

## A.5 An upper limit for the minimum 'L0 norm'

We use a known fact from graph theory: The number of (basic) cycles in a graph $G$ (which corresponds to the degrees of freedom) is given by $c = e - v + s$, where $e$ is the number of edges, $v$ is the number of vertices, and $s$ is the number of connected components (subgraphs) of $G$.[20,21] $c$ can be interpreted as the number of edges that have to be 'cut' to make the graph a tree (i.e., a graph free of cycles).

**Proposition 2.** The minimum $L0$ norm of $\phi$ does not exceed $c + 1$.

*Proof.* We assume a network graph $G = (V, E)$ and consider a particular comparison A:B. We want to list all non-zero elements of $\phi_{A:B}$ which correspond to streams from A (the source) to B (the sink). The flow from A to B is based on the row $\mathbf{h}_{A:B}$ of the hat matrix $\mathbf{H}$. Without loss of generality, we may assume that all elements of $\mathbf{h}_{A:B}$ are positive, otherwise we only need to change the direction of all comparisons that belong to a negative entry. We may ignore edges with zero flow. We recursively iterate two steps.

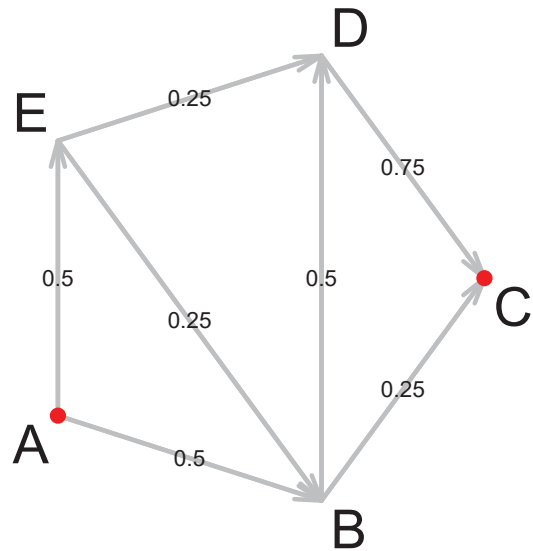

**Figure A1** Directed network graph for counterexample with hat matrix entries for comparison A:C.

1. (Step 1) We identify a path $p$ from A to B and subtract the flow along its edges by the minimum flow in the path, $h_p$. This results in taking away all edges with minimal flow. Note that we do not remove the vertices ($v' = v$). So we are left with a new graph $G' = (V, E')$ with $e' < e$. We have

$$c - c' = (e - v + s) - (e' - v + s') = e - e' - (s' - s)$$

and thus

$$(c - c') + (s' - s) = e - e' > 0$$

which means that at least one of the numbers $c - c'$ or $s' - s$ (or both) must be positive: Either the number of cycles is reduced by (at least) one, or the graph has been separated into further components (including possibly isolated vertices), or both.

2. (Step 2) Concerning the matrix row $\mathbf{h}_{A:B}$, we reduce all elements in $\mathbf{h}_{A:B}$ that correspond to a piece of path $p$ from Step 1 by the amount $h_p$. This provides a new vector $\mathbf{h}_p$, containing zeros at those places where edges were removed. $\mathbf{h}_p$ corresponds to a graph that is obtained from the starting graph by removing one or more edge(s) from path $p$ and accordingly reducing the flows. The element of $\phi_{A:B}$ which corresponds to path $p$ is now set to $h_p$.

Steps 1 and 2 are recursively repeated, at most $c$ times until there are no more cycles, at which point at most one path remains, which fully exhausts the flow. Thus we have found at most $c + 1$ paths (or less, if either several loops were broken in one step or the graph was disconnected before the maximal number of steps was reached). This proves the proposition. □

*Remark 4.* This proposition gives an upper limit for the number of positive path weights for *shortestpath*, but does not provide an algorithm for minimizing the 'L0 norm'.